# Enhanced Light-Matter Interaction in B-10 Monoisotopic Boron Nitride Infrared Nanoresonators

*Marta Autore, Irene Dolado, Peining Li, Ruben Esteban, Ainhoa Atxabal, Song Liu, James H. Edgar, Saül Vélez, Fèlix Casanova, Luis E. Hueso, Javier Aizpurua, and Rainer Hillenbrand[*]*

Dr. M. Autore, I. Dolado, Dr. P. Li, Dr. A. Atxabal, Prof. F. Casanova, Prof. L. E. Hueso, Prof. R. Hillenbrand
CIC nanoGUNE, 20018 Donostia-San Sebastián, Spain
Email: r.hillenbrand@nanogune.eu

Dr. R. Esteban, Prof. J. Aizpurua
Donostia International Physics Center (DIPC), Donostia-San Sebastián 20018, Spain

Dr. R. Esteban, Prof. F. Casanova, Prof. L. E. Hueso, Prof. R. Hillenbrand
IKERBASQUE, Basque Foundation for Science, Bilbao 48013, Spain

Dr. S. Liu, Prof. J. H. Edgar
Tim Taylor Department of Chemical Engineering, Kansas State University, Manhattan, KS 66506, USA

Dr. S. Vélez
Department of Materials, ETH Zürich, 8093 Zürich, Switzerland

Prof. J. Aizpurua
Centro de Física de Materiales (MPC, CSIC-UPV/EHU), 20018 Donostia-San Sebastián, Spain

Prof. R. Hillenbrand
EHU/UPV, 20018 Donostia-San Sebastián, Spain

**Keywords:** boron nitride, phonon-polaritons, strong coupling, surface enhanced infrared absorption.

Phonon-polaritons, mixed excitations of light coupled to lattice vibrations (phonons), are emerging as a powerful platform for nanophotonic applications. This is because of their ability to concentrate light into extreme sub-wavelength scales and because of their longer phonon lifetimes than their plasmonic counterparts.

In this work, the infrared properties of phonon-polaritonic nanoresonators made of monoisotopic B-10 hexagonal-boron nitride (h-BN) are explored, a material with increased phonon-polariton lifetimes compared to naturally abundant h-BN due to reduced photon scattering from randomly distributed isotopes. An average relative improvement of 50% in the nanoresonators Q factor is obtained with respect of nanoresonators made of naturally abundant h-BN.

Moreover, the monoisotopic h-BN nano-ribbon arrays are used to sense nanometric-thick films of molecules, both through surface-enhanced absorption spectroscopy and refractive index sensing. In addition, strong coupling is achieved between a molecular vibration and the phonon-polariton resonance in monoisotopic h-BN ribbons.

**1. Introduction**

The emerging field of nanophotonics requires a broad range of low-loss sub-diffractional components, for full light control and manipulation at the nanoscale, such as amplitude and phase modulators, waveguides, perfect absorbers, light generators, concentrators and switches. Most of the research during last decade has focused on plasmonic structures based on metals[1, 2] and, more recently, on high refractive index dielectrics[3] and graphene.[4]

Phonon-polaritonic nanostructures made of polar crystals (SiC, BN, Quartz…) represent a valuable platform for infrared nanophotonics due to the higher Q factor than their plasmonic counterparts and to their high degree of the electromagnetic field spatial confinement.[5,6,7,8,9,10] This is achieved through the excitation of surface phonon-polaritons, mixed excitations of photons and crystal lattice vibrations, which is possible in the so-called *reststrahlen bands*, in which the real part of the dielectric function is negative, between the transverse optical (TO) and the longitudinal optical (LO) phonons.

Recently, van der Waals stacks of materials have opened new opportunities towards fabricating nanophotonic components,[11] due to the possibility to exfoliate atomically flat, few layer-flakes of layered materials. A prototypical example is hexagonal-boron nitride (h-BN).[12] The layered nature of h-BN leads to the strong uniaxial anisotropy of its dielectric function in the in-plane ($\epsilon_\parallel$) and out-of-plane ($\epsilon_\perp$) direction. In its reststrahlen bands, where $\text{Re}(\epsilon_\parallel) \cdot \text{Re}(\epsilon_\perp) < 0$, the phonon-polaritons can propagate inside the volume of the material with hyperbolic dispersion (and therefore they are called hyperbolic phonon-polaritons, HPhPs) and show negative phase velocity and slow group velocity.[13] The great potential of h-BN for nanophotonics has been recently unveiled by fabrication of very high Q resonators,[14,15] linear waveguide antennas[16] and hyperbolic metasurfaces.[17] Moreover, h-BN nanoribbons have been used for surface-enhanced infrared absorption (SEIRA) spectroscopy[18,19] where their high Q factor is beneficial for approaching the strong coupling of the HPhPs to a molecular vibration.[15]

Very recently, Giles *et al.* substantially increased the propagation length and lifetime of phonon-polaritons in h-BN, by employing monoisotopic (isotopically pure) h-BN.[20] Indeed, while $^{14}$N is 99.6% abundant, natural boron consists of ~80% $^{11}$B and ~20% $^{10}$B isotopes. The random distribution of boron isotopes (isotopic disorder) leads to increased phonon scattering in natural h-BN. Giles *et al.* showed that h-BN crystals with high isotope purities 99.22 at% $^{10}$B and 99.41 at% $^{11}$B increased the lifetime of HPhPs propagating in h-BN thin flakes by factors between 1.4 and 1.8 depending on the frequency range. First-principle calculations predict even higher increases are possible, as much as an order of magnitude. However, the effect of this lifetime increase on the Q factor of phonon-polaritonic resonators, where fabrication-related issues also play an important role, is to our knowledge still to be studied.

In this work we fabricated nanoribbon arrays made of monoisotopic h-BN (99.22 at% $^{10}$B),[21] indicated as h-BN* in the following (see Methods for details on crystal growth). By measuring far-field transmission spectra, we show an increase of the Q factor of the HPhPs lowest-order Fabry-Perot resonance in h-BN* compared to natural abundant h-BN ribbon arrays, and the appearance of even higher Q-factor modes at lower frequencies. In the second part of the paper we show how h-BN* nanoribbons can be used for simultaneous refractive index sensing and SEIRA spectroscopy of thin molecular layers. Importantly, the improvement in the Q factor of the HPhP resonances leads to the full achievement of the strong coupling between phonon polaritons and a weak molecular vibration, which was not possible in recent experiments performed with naturally boron isotope abundant h-BN.[15]

## 2. Comparison between naturally abundant and monoisotopic h-BN nanoresonators

To compare the response of nanoresonators made of monoisotopic and naturally abundant h-BN, we fabricated several sets of ribbon arrays (with varying widths $w$ nominally ranging from 140 to 200 nm, and fixed period $D = 400$ nm) starting from mechanically exfoliated flakes of comparable thickness $d \approx 35$ nm. All the fabrication steps (mechanical flake exfoliation, electron beam lithography and chemical dry etching) were carried out at the same time and on the same substrate (a 10x10x1 mm$^3$ CaF$_2$ crystal), to avoid possible differences induced by changes in the fabrication process. Indeed, previous experiments showed that the experimental Q factor of h-BN nanoresonators is lower then the theoretical one[15] (see also the results of next Section), probably due to increased HPhP scattering induced by the fabrication process.

In **Figure 1** we show the transmission spectra of natural (panel a) and monoisotopic (panel b) h-BN ribbons, measured via Fourier transform infrared spectroscopy using a Vertex interferometer coupled to a Hyperion 2000 IR microscope (Bruker). All spectra were measured in an $N_2$-purged box at room temperature and normalized to the bare substrate spectrum, $T_0$. Normally incident light was polarized perpendicular to the ribbons, in order to excite the first order transversal Fabry-Perot resonance of the HPhP.[15]

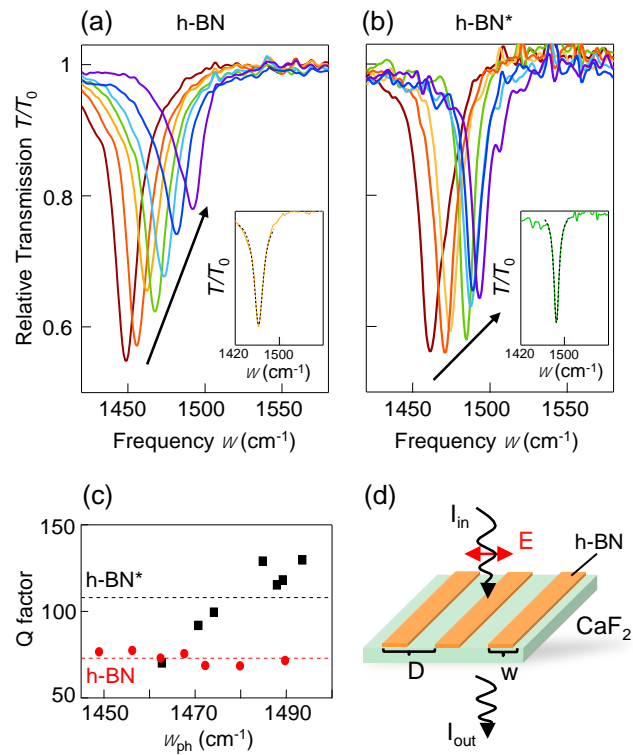

**Figure 1**. Relative transmission spectra of (a) naturally abundant h-BN and (b) monoisotopic h-BN ribbon arrays of varying $w$, measured at normal incidence and polarization perpendicular to the ribbons. Nominal values of $w$ ranging from 140 nm to 200 nm correspond to colors from purple to red (black arrow indicates decreasing $w$). Insets: exemplary transmission curve showing the Lorentzian fit (dashed line) to the HPhP resonance dip (solid line). (c) Calculated Q factor of each resonator, plotted as a function of the resonator central frequency for h-BN (red dots) and h-BN* (black squares). Dashed lines represent the respective average values. (d) Sketch of the FTIR transmission experiment.

The resonance appears as a pronounced dip in the curves and blueshifts as the width of the ribbon decreases. A quantitative analysis of the measured spectra allows us to quantify the difference in the quality factor of resonances for the two materials. To that end, we fit a Lorentzian lineshape (see Methods) to each curve in order to extract the resonances' central frequency $\omega_{\text{HPhP}}$ and full width at half maximum (FWHM), $\gamma_{\text{HPhP}}$. We obtain the quality factor $Q = \frac{\omega_{\text{HPhP}}}{\gamma_{\text{HPhP}}}$, which is shown in Figure 1c. There is an overall improvement of the quality factor of the h-BN* ribbons compared to that of the ribbons made of naturally abundant h-BN. However, the improvement is not constant over the considered frequency range. At higher frequencies ($\omega > 1480$ cm$^{-1}$), Q is consistently about 70% higher, but decreases with decreasing frequency. At 1460 cm$^{-1}$ the quality factor of h-BN and h-BN* ribbons are equal, suggesting that for this array fabrication-induced mechanisms of HPhP scattering might have played a relevant role. However, the average relative improvement (50%) of Q is consistent with the values of the lifetime increase reported by Giles et al.[20] for propagating HPhPs in a similar frequency range (1400 to 1500 cm$^{-1}$).

**3. Observation of higher order modes in thick monoisotopic h-BN nanoribbon arrays**

We also performed transmission spectroscopy on a set of thicker ribbon arrays ($d = 90$ nm), with fixed period $D = 400$ nm and several widths $w$. All spectra (**Figure 2**a) show a pronounced dip above 1500 cm$^{-1}$, denoted by the letter α in the plot, and two smaller dips (indicated as β and γ). The latter are superimposed to a much broader dip centered around 1400 cm$^{-1}$ and associated to residual contribution from the h-BN* TO phonon. Interestingly, the β and γ dips possess higher Q factors up to 310, although their intensity is lower than the fundamental (α) resonance (see Supplement for details). These high Q factors could be beneficial for mid-infrared applications that require extremely narrow bandwidth, such as narrowband thermal emitters.[22]

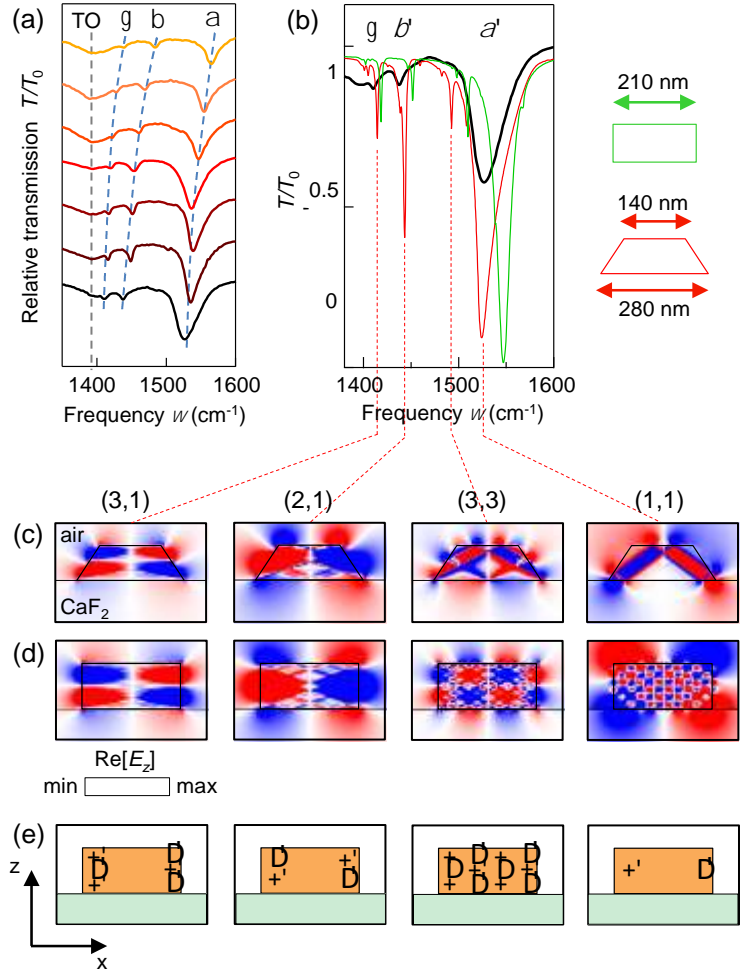

**Figure 2**. (a) Relative transmission spectra of 90 nm-thick h-BN* ribbon arrays of several widths $w$, (nominal $w$ ranging from 130 nm, yellow, to 190 nm, black) with light linearly polarized perpendicular to the ribbons. α, β and γ symbols indicate the three peaks corresponding to HPhP resonances. The curved blue dashed lines are guide to the eyes. The vertical grey dashed line indicates the TO phonon frequency. (b) Comparison between a measured spectrum (black), and the simulated spectrum for rectangular (green, $w$ =210 nm) and trapezoidal (red) ribbons with top (bottom) width of 140 nm (280 nm). Widths were measured by atomic force microscopy. (c-d) Simulated real part of the vertical component of the electric field, Re[$E_z$], for the peaks of order (3,1), (2,1), (3,3) and (1,1) (from left to right, respectively) for the case of trapezoidal (c) and rectangular (d) ribbons. (e) Effective charge profile of the observed modes for the rectangular ribbons case.

All the dips in Figure 2a shift to lower frequencies with increasing width (from yellow to black spectrum), as highlighted by the blue dashed guide to the eyes. This behavior suggests that they are associated to resonating HPhP higher order modes. The observation of higher-order modes localized resonances has been previously reported in h-BN nanostructures of thickness $d \geq 250$ nm (Refs. 14, 23, 24), more than a factor of 2 thicker than our case.

To better understand the nature of these resonances, in Figure 2b we compare an exemplary spectrum (black curve, ribbon arrays $w = 210$ nm, measured by AFM as the average between top and bottom surface) to a spectrum calculated via electromagnetic simulations (green curve, see Methods for details) for rectangular ribbons of the same width. The two curves present the same behavior, but the peaks of the simulated ribbons are all blue-shifted compared to the experimental ones. A better agreement is reached by simulating a more realistic structure, with ribbons of cross sections of trapezoidal shape (red), with bases of 140 and 280 nm, as measured by AFM. It is worth to note that the β and γ features in the simulations possess extremely high Q factors (420 and 680, respectively), and we probably cannot resolve them accurately in the experiment (with a spectral resolution of 2 cm$^{-1}$).

To elucidate the character of the modes associated to the resonance peaks, we calculated the electric field distribution at the peak positions. Cross section of Re[$E_z$] for ribbons with trapezoidal and rectangular shapes are shown in Figure 2c and 2d, respectively. The latter are useful to better understand the complex field pattern produced in the trapezoidal case. To identify the modes, we first analyse the field outside of the rectangular ribbons (neglecting the complicated pattern inside the structures, which is due to the superposition of hyperbolic rays[16]). The effective charge profile corresponding to the observed modes is sketched in Figure 2e, as a guide. We can clearly recognize the transversal structure of the modes in both the $z$ and $x$ direction. The number of nodes of Re[$E_z$] in the $z$-direction indicates the order, *j*, of the so-called slab modes (we follow the notation of Ref. 25). In the *x*-direction, we can instead recognize the order, *k*, of Fabry-Pérot resonances. Following this notation, we are able to identify and label with indices (*j*, *k*) the four main modes in the simulations and thus associate the experimentally observed three peaks α, β and γ to modes (1,1), (2,1) and (3,1), respectively. The mode (3,3), which is present in simulations, was not present in the experimental spectra.

In the case of the trapezoidal shaped ribbons (Figure 2c) the field structure is substantially maintained the same as for rectangular ribbons, with the same number of nodes in the *x* and *z* directions, but with a different spatial distribution. This comparison allows us to identify the actual observed modes in the trapezoidal ribbons in terms of first order Fabry-Pérot modes of higher-order slab modes, as the one identified for the simulated rectangular ones.

## 4. Simultaneous SEIRA and refractive index sensing with enriched h-BN resonators

h-BN* nanoresonators offer a great potential for sensing experiments that benefit both from the field enhancement on the resonators' surface and from the narrow linewidth of the resonances.[26] In particular, we show that they are promising for simultaneous surface-enhanced infrared absorption (SEIRA) spectroscopy[18] and refractive index (RI) sensing. The combination of the two sensing methodologies in the mid-IR has rarely been investigated up to now, despite interesting results.[27,28,29,30,31]

To explore the possibility of simultaneous SEIRA and mid-infrared RI sensing with h-BN* ribbon arrays, we evaporated layers of the organic semiconductor 4,4'-bis(N-carbazolyl)-1,1'-biphenyl (CBP)[32] on top of the arrays (see sketch in **Figure 3**d). In Figure 3a and 3b we report the spectra of two sets of arrays with different width *w* =

230 nm and 270 nm (indicating the average of the width of trapezoidal ribbons, as measured by AFM), respectively, with several thicknesses (1, 3, 5, 10, 30 and 50 nm) of CBP on top. The CBP molecule exhibits a vibrational mode at 1450 cm$^{-1}$ (corresponding to a dip in the spectrum of a 30 nm thick film, plotted as red curve), associated to a deformation of the C-H bond located on the carbazole rings.[32] From the transmission spectra, we can clearly observe the modifications induced by the presence of CBP on top of the ribbons. The effect of the CBP layer is due to a spectral feature associated to the enhanced molecular vibration and its coupling to the HPhP resonance, and an overall redshift of the spectra as the CBP thickness increases.

To artificially separate these two contributions, we fitted a two coupled oscillators model to the spectra of the *w* = 270 nm ribbons (Figure 3(b)). The model reproduces the emergence of the feature associated to the molecular vibration and the progressive splitting of the HPhP dip, as the CBP thickness increases, which characterizes the regimes of weak and strong coupling (which will be discussed in the next Section).

The model is presented in Ref. 15 and describes two classical oscillators (each one characterized by central frequency, linewidth and intensity) interacting via a coupling constant *g*. In the fitting, the central frequency and linewidth of the oscillator corresponding to the CBP are fixed (as taken from its dielectric function, see Ref. 15) and the fitting parameters are the ones relative to the HPhP and coupling strength. By setting *g* = 0, we can retrieve the shape of the uncoupled resonances. The resulting uncoupled HPhP spectra, $\frac{T_{\text{HPhP}}}{T_0}$, are plotted as orange lines in Figure 3b. Note that the model takes into account the non-dispersive part of the dielectric function of CBP ($\varepsilon_\infty = 2.8$) by setting the resonant HPhP frequency as a fitting parameter, which explains why the obtained uncoupled HPhP spectra are red-shifted in frequency compared to the measured spectrum of h-BN* ribbons with no CBP on top. In summary, the uncoupled HPhP spectra correspond to the phonon-polariton resonance spectra if they were affected just by the change of the dielectric environment due to $\varepsilon_\infty$, thus enabling us to study the potential of SEIRA and refractive index sensing separately, as we explain in the following.

To quantify the effect of SEIRA, we studied the quantity $\Delta T = \frac{T}{T_0} - \frac{T_{\text{HPhP}}}{T_0}$, i.e. the difference between the measured spectra of h-BN* ribbons with CBP on top (HPhP interacting with the molecular vibration) and the uncoupled HPhP spectra resulting from the fit. $\Delta T$ is plotted in panel (c) for each value of CBP thickness and highlights the strong enhancement of the CBP vibration signature due to the ribbons. For the 10 nm-thick bare CBP layer, the vibrational signature amounts to around 1% of the transmission (green line). Interestingly, the contrast becomes 15% in $\Delta T$, when the layer is placed on top of the ribbons (as indicated in Figure 3c). Analogously, for the case of 30 nm-thick CBP, the signature of 3% for bare CBP (red line) becomes 27% in $\Delta T$.

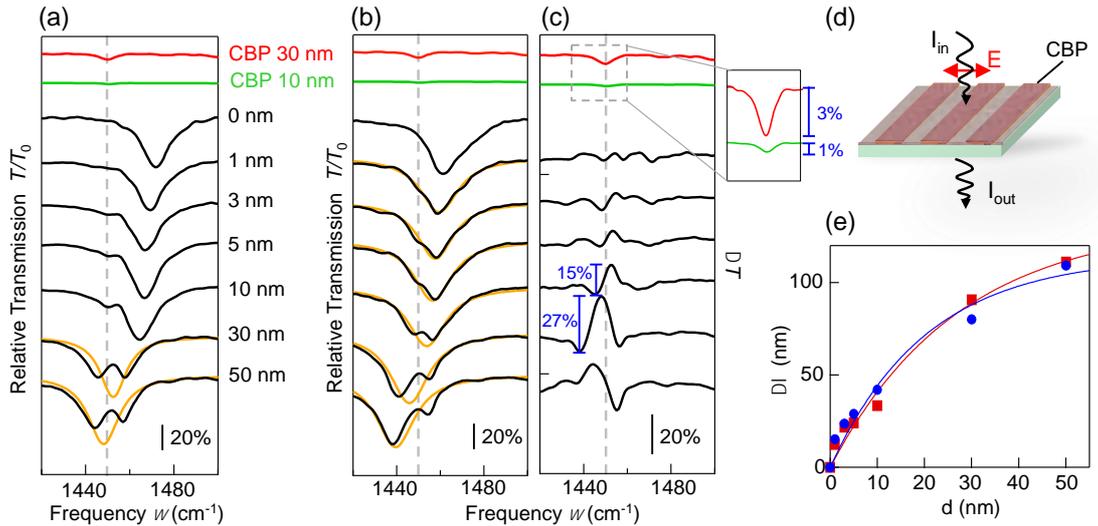

**Figure 3**. (a) Transmission spectra of a set of h-BN* ribbon array *(D* = 400 nm, average *w* = 230 nm)* with several thickness layers of CBP molecules on top (black curves), according to the legend. As a reference, the spectra of 30 nm and 10 nm-thick CBP layers are plotted in red and green, respectively. The grey dashed line marks the spectral position of the vibration associated with the C-H bond deformation. The orange curves show the uncoupled HPhP spectra, as extracted from the coupled oscillators model fit. (b) Same as panel (a), but for and array of ribbons with average *w* = 270 nm. (c)

Enhancement signal $\Delta T$, for the array of panel (b). The inset shows a zoom of the spectra of bare CBP in order to evaluate the magnitude of the vibration signature. Respective scalebars are indicated in panel (a-c). (d) Sketch of the transmission experiment. (e) Shift of the HPhP resonance position $\Delta\lambda$ due to dielectric environment change, as a function of the CBP layer thickness for the set of array of $w$ = 230 nm (red squares) and $w$ = 270 nm. Fits of Eq. 1 to data are plotted as continuous lines.

Having proved the potential of h-BN* ribbons for SEIRA spectroscopy, we explore the possibility of exploiting HPhP resonances for refractive index sensing, i.e. by quantifying the resonance shift as a function of the layer thickness. To do so, we consider the position of the uncoupled HPhP resonances, since this enables us to isolate the effect of the change in the non-dispersive dielectric background due to the presence of the molecular layer.

The shift of the HPhP dip from its original position (in the case of no CBP on top of ribbons), $\Delta\lambda$, was measured as a function of the CBP layer thickness $d$. We perform this analysis in wavelength units, after a conversion from frequency units, in order to better compare the results with literature. $\Delta\lambda$ for both the set of h-BN ribbons with $w$=230 nm and $w$=270 nm is plotted in Figure 3e as red squares and blue dots, respectively.

While the shift is evident down to 1 nm-thick CBP layer, quantitative evaluation of the result requires some considerations. The conventional figure of merit (FOM) for RI sensing is given by $\text{FOM}_{\text{bulk}} = \partial\lambda/\partial n$, describing the change of the wavelength of a (conventionally) plasmonic resonance as the surrounding refractive index, $n$, is varied. However, it has been recently argued that this quantity does not represent the most significant quantity when evaluating the performance of a biosensor,[33,34] since in this situation the analyte is a thin (usually nanometric-thick) film placed on the sensor surface, implying that the dielectric environment changes only in its close proximity. In this case, it is more appropriate to evaluate the surface sensitivity, which depends on the specific electric field distribution on top of the nanostructured resonating substrate and strongly depends on its decay length. In recent experiments using plasmonic nanostructures[33,34], it has been proposed to quantify the plasmonic spectral shift $\Delta\lambda$ as a function of the thickness $t$ of the analyte as follows:

$$\Delta\lambda = m\,\Delta n\left(1 - e^{-\frac{2t}{l_d}}\right) \qquad (1)$$

where $\Delta n$ is the variation of the refractive index due to the analyte and $l_d$ is the electric field decay length in a simplified scenario with a 1D field distribution along the vertical axis. In a more realistic scenario, the field distribution is highly inhomogeneous in the 3D space and $l_d$ will assume the meaning of an *effective* decay length. The parameter $m$ corresponds to the bulk FOM for the case $t \gg l_d$. In this framework, the so-called second-order surface sensitivity can be calculated as the second order derivative of the spectral shift $\Delta\lambda$,

$$\frac{\partial^2 \Delta\lambda}{\partial n\, \partial t} = \frac{2m}{l_d} e^{-\frac{2t}{l_d}} \qquad (2)$$

The prefactor of this expression, $\frac{2m}{l_d}$, meaning the extrapolation of the sensitivity for analyte thickness approaching to zero, can be used to compare the performance of different biosensors.[33,34,35] We fit Eq.1 to the extracted $\Delta\lambda$ values, setting $\Delta n = 0.67$ according to the CBP non-dispersive dielectric background $\varepsilon_\infty = 2.8$ (Refs. 15, 32). The resulting fits are plotted as thin lines in Figure 3e. The extracted parameters are $m$ = 198 nm/RIU and $l_d$ = 54 nm for the $w$=230 nm ribbon array, and $m$ = 171 nm/RIU and $l_d$ = 41 nm for the $w$=270 nm ribbon array. These values yield $\frac{2m}{l_d}$ = 7.3 RIU$^{-1}$ and $\frac{2m}{l_d}$ = 8.3 RIU$^{-1}$, respectively. These values are higher or similar to the ones reported for diffractively coupled plasmonic crystals in the visible range,[33,35] revealing the great potential of h-BN* ribbon arrays for mid IR refractive index sensing.

**5. Vibrational strong coupling with enriched h-BN resonators**

In our recent work,[15] we showed that using nanoresonators made of naturally abundant h-BN we were able to approach the limit of strong coupling between the HPhP resonance and the CBP molecular vibration. In the strong coupling regime the original states mix, forming two new hybridized polaritonic states.

The higher Q obtained using h-BN* could represent a decisive element towards the achievement of vibrational strong coupling. To verify this hypothesis, we repeat the experiment with a set of h-BN* ribbon arrays. We evaporated 30 nm of CBP on top of the arrays and measured IR transmission spectra (**Figure 4**a). All spectra show a double-dip structure, exhibiting an anti-crossing behavior when approaching the CBP vibrational mode at 1450 cm$^{-1}$. To evaluate the presence of strong coupling, the mere observation of anti-crossing in the transmission spectra is not sufficient, since it does not always imply a hybridization of the interacting modes into two new eigenstates. For a proper evaluation of the coupling regime, it is necessary to retrieve the value of the coupling constant between the two modes, *g*, and to compare it to the losses of the system. This can be done by fitting to each spectrum the classical model for two coupled oscillators used in previous sections, each one described in terms of a central frequency $\omega_i$, a width $\gamma_i$, and an effective driving force $F_i$ (proportional to the external electromagnetic field) and coupled to each other through a constant *g*. A complete description of the model is described in Ref. 15.

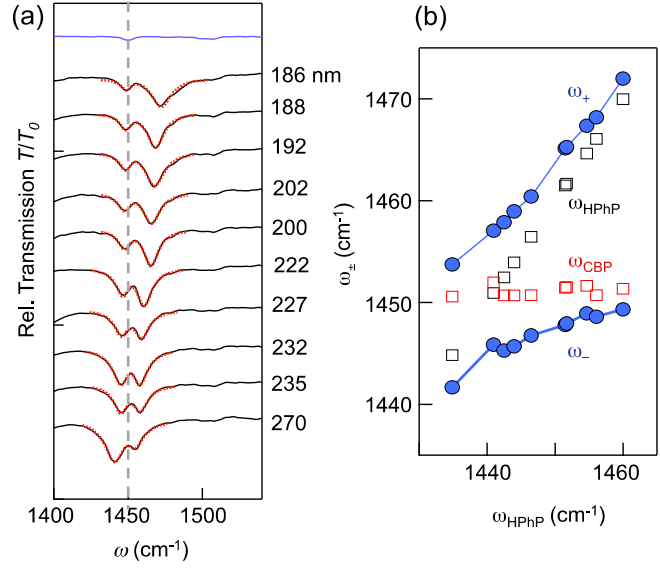

**Figure 4**. (a) Relative transmission spectra of 32 nm-thick h-BN* ribbon arrays with 30 nm CBP on top (black curves. The ribbon width in nm is indicated on graph, measured by AFM as the average between the top surface and bottom base). Red curves are the fit according to the coupled oscillators model. Blue curve represents the spectrum of 30 nm CBP directly on the CaF2 substrate. (b) Eigenstate frequencies $\omega_\pm$ (blue dots) as calculated with the parameters obtained from the fits in panel (a). The bare CBP vibration frequency ($\omega_{CBP}$) and bare HPhP resonance frequency ($\omega_{HPhP}$) obtained for each fit are reported as red and black squares, respectively.

Fits are plotted as red dotted lines in Figure 4a and from each array the coupling strength $g$, the bare HPhP frequency $\omega_{HPhP}$ and width $\gamma_{HPhP}$ were retrieved (the bare molecular vibration frequency $\omega_{CBP}$ and width $\gamma_{CBP}$ were fixed according to the CBP dielectric function, as in previous Section). These parameters can be used to calculate the energies of the new eigenstates of the coupled system,[36] $\omega_\pm$, plotted as blue dots in Figure. 4b as a function of $\omega_{HPhP}$ (see Supplementary Information for details). The anti-crossing of the $\omega_\pm$ branches is a necessary condition for strong coupling and corresponds to the mathematical condition $C_1 \stackrel{\text{def}}{=} \frac{|g|}{|\gamma_{HPhP}-\gamma_{CBP}|} > 0.25$ (refs 37, 38). Using the average values extracted from the fits ($g$ = 6.4 cm$^{-1}$, $\gamma_{HPhP}$ = 15.0 cm$^{-1}$, $\gamma_{CBP}$ = 8.5 cm$^{-1}$) we obtain $C_1 \cong 1.0$, indicating strong coupling. However, a more restrictive criterion to indicate the threshold to practically observe the effects of strong coupling is given by $C_2 \stackrel{\text{def}}{=} |g|/|\gamma_{HPhP}+\gamma_{CBP}| \gtrsim 0.25$ (ref 37). Here, we obtain $C_2 \cong 0.27$, thus fulfilling even the more strict condition. This second condition

was not fulfilled in the case of naturally abundant h-BN, as reported in Ref. 15. An alternative condition to verify strong coupling, also found in literature,[37] is based on verifying that the normal mode splitting at zero detuning is larger than the losses of the system, and can be expressed as $\omega_+ - \omega_- > \frac{\gamma_{\text{HPhP}}}{2} + \frac{\gamma_{\text{CBP}}}{2}$. This condition is fulfilled in our experiment, in which we obtain 12.6 cm$^{-1}$>11.5 cm$^{-1}$.

We corroborate our finding with the help of electromagnetic simulations, in which we verify the splitting in the calculated absorption of the CBP molecules when they are placed on top of the h-BN ribbons (see Supplementary Information), a further signature of the mixed state character of a strongly coupled system.[39]

We verify that the high Q of h-BN* ribbons is crucial to achieve strong coupling. Indeed, the average Q calculated using the parameters extracted from the fits in Figure 4a is $Q_{\text{ave}}$= 98 (see Supplementary Information for details), similar to the value reported in Figure 1c. For this comparison, note that the measurement in Figure 1 were performed without molecules present, so that the resonances would redshift if we deposited the same CBP layer as in Figure 4. By using the average Q factor reported in Figure 1c for natural abundance h-BN ribbons, we would not fulfill the conditions $C_2 > 0.25$ and $\omega_+ - \omega_- > \frac{\gamma_{\text{HPhP}}}{2} + \frac{\gamma_{\text{CBP}}}{2}$.

## 6. Conclusion

In summary, we have investigated the infrared response of phonon-polaritonic nanoribbon arrays of monoisotopic B-10 h-BN and their potential for sensing and strong coupling applications. Compared to ribbon arrays made of naturally abundant h-BN and fabricated in the same process, the monoisotopic h-BN had higher Q factor over most of the explored frequency range, suggesting better performances in applications requiring narrow-band resonances. Moreover, we showed that even higher Q factors can be obtained for higher-order resonances (up to Q = 310) in thicker nanoresonators ($d$ = 90 nm).

Further, we studied the performance of monoisotopic h-BN nanoribbons for simultaneous SEIRA spectroscopy and refractive index sensing. To that end, we deposited CBP layers of varying thickness onto the ribbons and we showed SEIRA enhancement of the CBP vibration signature reaching more than a factor of 10 compared to the signature detected in bare CBP spectra of comparable thickness. Regarding the refractive index sensing capabilities, we evaluated the surface sensitivity of our devices, the most relevant quantity to be considered when dealing with thin films of analyte. Our surface sensitivity was better or comparable to the ones reported for diffractively-coupled plasmonic crystals in the visible range.[33,35]

The combined SEIRA and refractive index sensing results show the high potential of monoisotopic h-BN nanostructures as a sensing platform, which could be further explored by optimizing the geometry of the resonators or by introducing some mechanism to further enhance the excitation of HPhP modes, such as integrating a metallic grating or using a Salisbury screen.[40,41]

Finally, we showed that the improved Q factor of monoisotopic h-BN resonators enables to fully reach the strong coupling regime between the molecular vibration and the HPhP resonances, which was not possible with naturally abundant h-BN nanoribbons. This result is an ultimate proof that monoisotopic h-BN nanoresonators represent a valid platform for further theoretical investigations and applications based on vibrational strong coupling. In particular, we envision the possibility to modify the chemical landscape[42, 43] and the speed of chemical reactions[44] at the nanoscale in a controlled way, by opportunely shaping and tuning the h-BN resonators, opening new opportunities in the field of nanoscale selective catalysis and quantum optical chemistry.[45]

## 7. Experimental Section

*Monoisotopic h-BN crystal growth method.* The $^{10}$B enriched h-BN crystals were grown from metal flux method. A Ni-Cr-$^{10}$B powder mixture at respective 48 wt%, 48 wt%, and 4 wt% was first loaded into an alumina crucible and placed in a single-zone furnace. The furnace was evacuated and then filled with $N_2$ and $H_2$ gases to a constant pressure of 850 Torr. During the reaction process, the $N_2$ and $H_2$ gases continuously flowed through the system at rates of 125 sccm and 5 sccm, respectively. All the nitrogen in the hBN crystal originated from the flowing $N_2$ gas. $H_2$ gas was used to minimize oxygen and carbon impurities in the hBN crystal. After a dwell time of 24 hours at 1550 °C, the hBN crystals were precipitated on the metal surface by cooling at a rate of 0.5 °C /h to 1525 °C, and then the system was quickly quenched to room temperature.

*Fabrication of h-BN ribbon arrays.* Large and homogeneous h-BN flakes were isolated and deposited on a $CaF_2$ substrate. To that end, we first performed mechanical exfoliation of commercially available h-BN crystals (HQ graphene Co, N2A1) using blue Nitto tape (Nitto Europe NV, Genk, Belgium). We then performed a second exfoliation of the h-BN flakes from the tape onto a transparent polydimethylsiloxane stamp. Using optical inspection and atomic force microscope (AFM) characterization of the h-BN flakes on the stamp, we identified high-quality flakes with large areas and appropriate thickness. These flakes were transferred onto a

CaF$_2$ substrate using the deterministic dry transfer technique.[46]

h-BN nanoribbon arrays of different widths were fabricated from the h-BN flakes by high-resolution electron beam lithography using poly(methyl methacrylate) (PMMA) as a positive resist and subsequent chemical dry etching. First, PMMA resist was spin coated over the substrate. Second, arrays of ribbons of different widths were patterned on top of the flakes (i.e., the inverse of the final h-BN ribbon array pattern was exposed to the electron beam). Third, the sample was developed in MIBK:IPA (3:1), leaving PMMA ribbons of the desired width and length on top of the flakes. Fourth, using these ribbons as a mask, the uncovered h-BN areas were chemically etched in a RIE Oxford Plasmalab 80 Plus reactive ion etcher (Oxford Instruments Plasma Technology, Bristol, UK) in a SF6/Ar 1:1 plasma mixture at 20 sccm flow, 100 mTorr pressure and 100 W power for 20 s. Finally, the PMMA mask was removed by immersing the sample overnight in acetone, rinsing it in IPA and drying it using a N$_2$ gun.

*Thermal evaporation of CBP.* 4,4′-bis(N-carbazolyl)-1,1′-biphenyl with sublimed quality (99.9%) (Sigma Aldrich) was thermally evaporated in an ultra-high vacuum evaporator chamber (base pressure < 10$^{-9}$ mbar), at a rate of 0.1 nm·s$^{-1}$ using a Knudsen cell.

*Fourier transform infrared (FTIR) micro-spectroscopy measurements.* Transmission spectra of the bare and molecule-coated h-BN arrays were recorded with a Bruker Hyperion 2000 infrared microscope coupled to a Vertex 70 FTIR spectrometer (Bruker Optics GmbH, Germany). The normal-incidence IR radiation from a thermal source (globar) was linearly polarized via a wire grid polarizer. The spectral resolution was 2 cm$^{-1}$. All measurements were taken at room temperature and ambient pressure, in a N$_2$-purged box.

*Fitting procedures for uncoupled and coupled HPhP resonances.* We fitted a Lorentzian lineshape to each spectrum in Figure 1a,b in order to extract their central frequency $\omega_{\text{HPhP}}$ and the FWHM $\gamma_{\text{HPhP}}$. We used the formula:

$$\frac{T}{T_0} = 1 - y_0 - \frac{A}{(\omega - \omega_{\text{HPhP}})^2 + \left(\frac{\gamma_{\text{HPhP}}}{2}\right)^2}$$

where $y_0$ indicates a small offset that might arise due to small variations in the experimental conditions, such as source emission oscillations.

In order to fit the spectra of the coupled HPhP-CBP vibration, we used the model described in detail in our previous work.[15]

The complete sets of parameters obtained by fitting are reported in the Supplementary Material.

*Electromagnetic simulations.* Full-wave numerical simulations using the finite-elements method in frequency domain (COMSOL) were performed to study the spectral response of h-BN* ribbons on a CaF$_2$ substrate. CaF$_2$ was described by a constant $\varepsilon_{\text{CaF2}} = 1.882$, while h-BN* dielectric permittivity was modeled according to $\varepsilon_a^{\text{h-BN}} = \varepsilon_{a,\infty}\left(1 + \frac{(\omega_a^{\text{LO}})^2 - (\omega_a^{\text{TO}})^2}{(\omega_a^{\text{TO}})^2 - \omega^2 - i\omega\gamma_a}\right)$, where $a = \parallel, \perp$ indicates the component parallel or perpendicular to the anisotropy axis. We use the parameters $\varepsilon_{\parallel,\infty} = 2.8$, $\omega_{\parallel}^{\text{TO}} = 785$ cm$^{-1}$, $\omega_{\parallel}^{\text{LO}} = 845$ cm$^{-1}$, $\gamma_{\parallel} = 1$ cm$^{-1}$, $\varepsilon_{\perp,\infty} = 3.0$, $\omega_{\perp}^{\text{TO}} = 1395$ cm$^{-1}$, $\omega_{\perp}^{\text{LO}} = 1630$ cm$^{-1}$, and $\gamma_{\perp} = 2$ cm$^{-1}$ (refs. 17, 20).

**Acknowledgements**

The authors acknowledge the European Union's Horizon 2020 research and innovation programme under grant agreement No GrapheneCore2 785219, the Spanish Ministry of Economy and Competitiveness under the Maria de Maeztu Units of Excellence Programme (MDM-2016-0618) and under projects MAT2015-65525, FIS2016-80174-P, MAT2015-65159-R and MAT2017-82071-ERC, the University of the Baque Country inder the project GIU18/202 and the Department of Economical Development and Infrastructures of the Basque Country inder the project Ekartek KK-2018/00001. The hBN crystal growth was supported by the National Science Foundation, grant CMMI 1538127 and the II-VI Foundation. A.A. acknowledges the Basque Government for a PhD fellowship (PRE_2017_2_0052).

**Competing financial interests**

The authors declare no competing financial interests.

*Marta Autore, Irene Dolado, Peining Li, Ruben Esteban, Ainhoa Atxabal, Song Liu, James H. Edgar, Saül Vélez, Fèlix Casanova, Luis E. Hueso, Javier Aizpurua, and Rainer Hillenbrand*[*]

Dr. M. Autore, I. Dolado, Dr. P. Li, Dr. A. Atxabal, Prof. F. Casanova, Prof. L. E. Hueso, Prof. R. Hillenbrand
CIC nanoGUNE, 20018 Donostia-San Sebastián, Spain
Email: r.hillenbrand@nanogune.eu

Dr. R. Esteban, Prof. J. Aizpurua
Donostia International Physics Center (DIPC), Donostia-San Sebastián 20018, Spain

Dr. R. Esteban, Prof. F. Casanova, Prof. L. E. Hueso, Prof. R. Hillenbrand
IKERBASQUE, Basque Foundation for Science, Bilbao 48013, Spain

Dr. S. Liu, Prof. J. H. Edgar
Tim Taylor Department of Chemical Engineering, Kansas State University, Manhattan, KS 66506, USA

Dr. S. Vélez
Department of Materials, ETH Zürich, 8093 Zürich, Switzerland

Prof. J. Aizpurua
Centro de Física de Materiales (MPC, CSIC-UPV/EHU), 20018 Donostia-San Sebastián, Spain

Prof. R. Hillenbrand
EHU/UPV, 20018 Donostia-San Sebastián, Spain

**S1. Lorentzian fits of transmission of h-BN and h-BN* ribbon arrays**

We fitted a Lorentzian lineshape to each spectrum in Fig. 1a,b of the main text, according to the formula reported in the Methods of the main text. In the following we show the fitting results (Fig. S1) for both h-BN (a) and h-BN* (b) sets of ribbon arrays. The obtained parameters are reported in Table T1.a and T1.b, respectively.

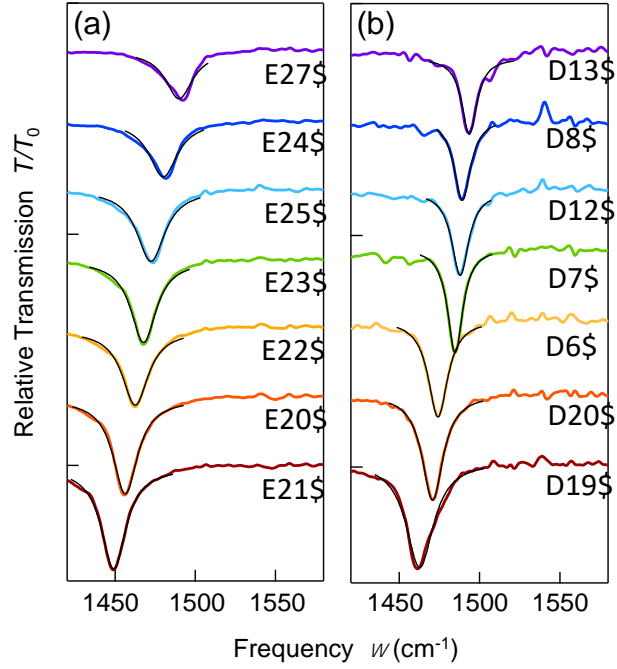

**Figure F1.** Relative transmission spectra of (a) h-BN and (b) h-BN* ribbon arrays of several widths. Black thin curves represent fits of Lorentzian lineshapes (Eq. 1) to data.

|  | $\omega_{HPhP}$ (cm$^{-1}$) | $\gamma_{HPhP}$ (cm$^{-1}$) | $A$ (cm$^{-2}$) | $y_0$ | $w$ (nm) |
|---|---|---|---|---|---|
| **E21** | 1449 | 18.9 | 38.9 | 0.012 | 200 |
| **E20** | 1456 | 18.8 | 35.9 | 0.016 | 190 |
| **E22** | 1462 | 20.0 | 32.1 | 0.018 | 180 |
| **E23** | 1468 | 19.4 | 33.1 | 0.016 | 170 |
| **E25** | 1472 | 21.4 | 35.3 | 0.009 | 160 |
| **E24** | 1480 | 21.6 | 27.9 | 0.012 | 150 |
| **E27** | 1490 | 20.8 | 22.1 | 0.009 | 140 |

**Table T1a.** Fitting parameters for Eq.1 to h-BN ribbons relative transmission spectra.

|  | $\omega_{HPhP}$ (cm$^{-1}$) | $\gamma_{HPhP}$ (cm$^{-1}$) | $A$ (cm$^{-2}$) | $y_0$ | $w$ (nm) |
|---|---|---|---|---|---|
| **D19** | 1463 | 20.8 | 49.4 | -0.023 | 200 |
| **D20** | 1471 | 16.0 | 29.1 | -0.013 | 190 |
| **D6** | 1474 | 14.8 | 22.9 | -0.035 | 190 |
| **D7** | 1485 | 11.5 | 14.9 | -0.044 | 180 |
| **D12** | 1488 | 12.9 | 15.0 | 0.012 | 150 |
| **D8** | 1489 | 12.6 | 14.0 | -0.005 | 170 |
| **D13** | 1494 | 11.5 | 10.9 | 0.034 | 140 |

**Table T1a.** Fitting parameters for Eq.1 to h-BN* ribbons relative transmission spectra.

## 2. Lorentzian fits of transmission of thick h-BN* ribbons with higher order modes

In order to fit the transmission spectra of ribbons made of a thick h-BN* flake (90 nm thick) we used the same model presented in section S1, considering the sum of 4 Lorentzian resonances, for the **(1,1), (2,1), (3,1)** modes described in the main text and for the residual contribution of the TO phonon of h-BN*, which is present because of eventual contribution of residual component of light polarized along the ribbons. We report the fit results (green thin lines) and the obtained Q factors for the **(1,1), (2,1), (3,1)** modes in Fig. F2 and Table T2, respectively.

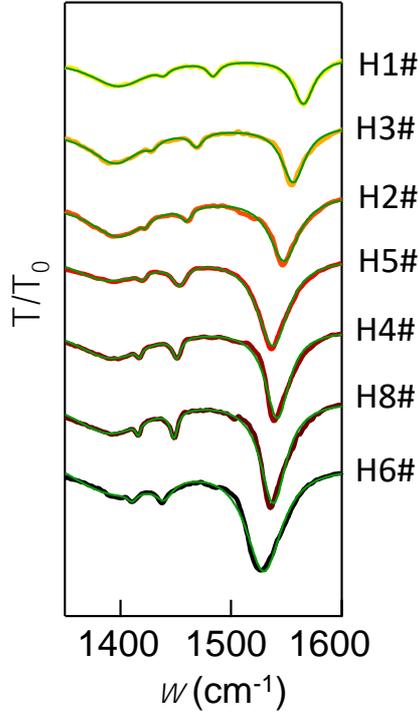

**Figure F2**. Fit to higher order modes.

| | H6 | H8 | H4 | H5 | H2 | H3 | H1 |
|---|---|---|---|---|---|---|---|
| **(1,1)** | 39 | 54 | 59 | 45 | 48 | 61 | 69 |
| **(2,1)** | 79 | 180 | 161 | 94 | 122 | 109 | 113 |
| **(3,1)** | 148 | 310 | 300 | 175 | 263 | 112 | 120 |

**Table T2**. Q factors extracted from fit to higher order modes.

## 3. Fits of coupled oscillator model for Fig. 4 (a) of the main text

To fit the transmission data of enriched h-BN* ribbons with 30 nm-thick CBP layer on top, we used a coupled oscillator model, previously presented in Ref. [1]. In the model, two classical oscillators represent the HPhP resonance and the molecular vibration, respectively. The equations of motion of the oscillators are given by

$$\ddot{x}_{HPhP}(t) + \gamma_{HPhP}\dot{x}_{HPhP}(t) + \omega^2_{HPhP}x_{HPhP}(t) - 2g\bar{\omega}x_{CBP}(t) = F_{HPhP}(t)$$

$$\ddot{x}_{CBP}(t) + \gamma_{CBP}\dot{x}_{CBP}(t) + \omega^2_{CBP}x_{CBP}(t) - 2g\bar{\omega}x_{HPhP}(t) = F_{CBP}(t)$$

where $x_{HPhP}$, $\omega_{HPhP}$, $\gamma_{HPhP}$ represent the displacement, frequency and damping of the HPhP oscillator, respectively. $F_{HPhP}$ represents the effective force that drives its motion and is proportional to the external electromagnetic field and $\bar{\omega} = \frac{\omega_{CBP}+\omega_{CBP}}{2}$. The corresponding notation is valid for the CBP vibration. The extinction ($E$) of such system can be calculated according to $E \propto \langle F_{HPhP}(t)\dot{x}_{HPhP}(t) + F_{CBP}(t)\dot{x}_{CBP}(t)\rangle$.

The transmission is related to the extinction by the relation: $\frac{T}{T_0} = 1 - E$. We added a constant parameter called *Offset* to this equation, to correct for some small mismatch of the signal, due, for instance, to thermal oscillations of the IR-light source.

The results of the fits are reported in Fig. 4a of the main text. Here we report the full set of values obtained from the fits for each sample. From the obtained parameters, we calculated the energy of the eigenstates of the coupled system $\omega_\pm$, reported in Fig 4b of the main text, and the average parameters $g = 6.4$ cm$^{-1}$, $\gamma_{HPhP} = 15.0$ cm$^{-1}$ (Q = 98.4), $\gamma_{CBP} = 8.5$ cm$^{-1}$, also reported in the main text.

The energy of the eigenstates of the coupled system $\omega_\pm$ was calculated as

$$\omega_\pm = \frac{1}{2}(\omega_{HPhP} + \omega_{CBP}) \pm \frac{1}{2}\sqrt{4|g|^2 + \left[\delta + i\left(\frac{\gamma_{CBP}}{2} - \frac{\gamma_{HPhP}}{2}\right)\right]^2}$$

as reported in Ref. 1, where we made the approximation $\omega - \omega_i \ll \omega_i$ and therefore $\omega^2 - \omega_i^2 \cong 2\omega_i(\omega - \omega_i)$, with i =HPhP, CBP.

| Sample | $\omega_{HPhP}$ | $\gamma_{HPhP}$ | F$_{HPhP}$ | $\omega_{CBP}$ | $\gamma_{CBP}$ | F$_{CBP}$ | Offset | g |
|---|---|---|---|---|---|---|---|---|
| D19 | 1444.8 | 18.4 | 3.3 | 1450.6 | 9.8 | -0.06 | -0.04 | 5.6 |
| D20 | 1451.0 | 15.7 | 3.0 | 1452 | 8.4 | -0.1 | -0.04 | 5.9 |
| D5 | 1452.5 | 13.2 | 2.6 | 1450.7 | 9.8 | 0.2 | -0.03 | 6.3 |
| D6 | 1453.9 | 12.9 | 2.5 | 1450.7 | 9.8 | 0.2 | -0.03 | 6.5 |
| D21 | 1456.5 | 13.7 | 2.8 | 1450.7 | 8.4 | 0.2 | -0.05 | 6.3 |
| **D7**  | 1461.5 | 12.9 | 2.6 | 1451.5 | 8.5 | 0.2 | -0.03 | 7.1 |
| **D26** | 1461.7 | 14.9 | 2.6 | 1451.6 | 8.4 | 0.2 | -0.05 | 7.1 |
| **D12** | 1464.6 | 17.2 | 2.8 | 1451.7 | 7.1 | 0.2 | -0.04 | 6.8 |
| **D8**  | 1466.1 | 14.2 | 2.6 | 1450.7 | 8.3 | 0.4 | -0.05 | 6.2 |
| **D13** | 1470.0 | 17.0 | 2.5 | 1451.3 | 6.8 | 0.2 | -0.03 | 6.6 |

**Table T3.** Parameters of the coupled oscillators model to fit the simulated transmission spectra. All parameters are expressed in cm$^{-1}$, except for *Offset* (adimensional) and $F_{HPhP}$, $F_{CBP}$ which are expressed in cm$^{-2}$.

### 4. Verification of strong coupling in calculated spectra

A way to verify that h-BN* ribbon arrays and CBP vibration are strongly coupled, is to observe a splitting in the absorption spectrum of the molecules only [2,3]. Unfortunately, this is not achievable in our experimental setup. However, we can calculate it with the help of electromagnetic simulations (see Methods in the main text). The CBP is described by the permittivity $\varepsilon_{CBP} = \varepsilon_\infty + \sum_k \frac{S_k^2}{\omega_k^2 - \omega^2 - i\omega\gamma_k}$ ($\varepsilon_\infty = 2.8$, $\omega_{CBP} = 1450.0$ cm$^{-1}$ and $\gamma_{CBP} = 8.3$ cm$^{-1}$ for the central frequency and width of the C-H deformation bond, respectively, which is the CBP vibration coupled to the HPhP resonance in this work. The values for the other vibrational modes in a close frequency range are reported in our previous work)[1].

The simulated transmission spectrum for sample D20 with 30 nm of CBP on top is shown in Fig. F4 (blue curve), together with the measured spectrum (blue dashed curve). In the same plot, we show the absorption of the CBP molecules (integrated over the whole volume) on top of ribbons (green) and bare substrate (black). The comparison clearly shows enhanced absorption and a splitting of the peak corresponding to the C-H vibration, which corroborates the presence of strong coupling.

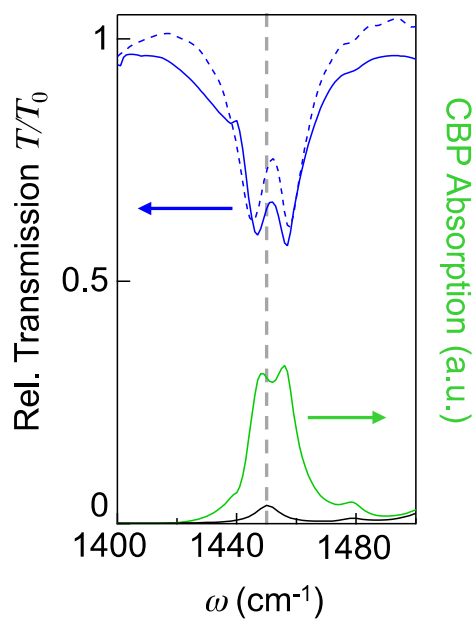

**Figure F4**. Simulated transmission spectrum for h-BN* ribbon arrays (top width of 190 nm, bottom 300 nm, period of 400 nm and thickness of 28nm) with a 30 nm-thick CBP layer on top (blue line), together with the experimental spectrum for the same sample (dashed blue line). Note that the measured experimental value for the ribbon thickness is 32 nm, and the value 28 nm was used in the simulations for better matching. Calculated absorption in the CBP molecular layer (integrated over the whole layer thickness) on top of the ribbons (substrate) is plotted as a green (black) curve. The vertical grey dashed line marks the molecular C-H vibration frequency for the uncoupled case.